\documentstyle[twocolumn,prl,aps,epsf]{revtex}

\catcode`@=11
\def\citer{\@ifnextchar [{\@tempswatrue\@citexr}{\@tempswafalse\@citexr[]}}
 
\def\@citexr[#1]#2{\if@filesw\immediate\write\@auxout{\string\citation{#2}}\fi
  \def\@citea{}\@cite{\@for\@citeb:=#2\do
    {\@citea\def\@citea{--\penalty\@m}\@ifundefined
       {b@\@citeb}{{\bf ?}\@warning
       {Citation `\@citeb' on page \thepage \space undefined}}%
\hbox{\csname b@\@citeb\endcsname}}}{#1}}
\catcode`@=12

\def\beq{\begin{equation}}
\def\eeq{\end{equation}}

\def\beqn{\begin{eqnarray}}
\def\eeqn{\end{eqnarray}}
\relax
\def\ba{\begin{array}}
\def\ea{\end{array}}
\def\squ{\tilde{Q}}
\def\tgb{\mbox{tg$\beta$}}
\def\hH{{\cal H}}

\begin{document}                                                              
\title{QCD Corrections to SUSY Higgs Production:
The Role of Squark Loops}  
\author{S.~Dawson$^1$, A.~Djouadi$^{2}$ and M.~Spira$^3$}
\address{
$^1$ Department of Physics, Brookhaven National Laboratory,
Upton, New York 11973-5000, USA \\
$^2$ Institut f\"ur Theoretische Physik, Universit\"at Karlsruhe, D--76128
Karlsruhe, Germany \\
$^3$ CERN, TH Division, CH--1211 Geneva 23, Switzerland}
\date{\today}
\maketitle
\begin{abstract}
We calculate the two--loop QCD corrections to the production of the
neutral supersymmetric Higgs bosons via the gluon fusion mechanism  at
hadron colliders, including the contributions of squark loops. To a good
approximation, these additional contributions lead to the same QCD
corrections as in the case where only top and bottom quark loops are
taken into account. The QCD corrections are large and increase the Higgs
production cross sections significantly. 
\end{abstract}



\begin{narrowtext}

The search for Higgs particles is an important component of the
experimental program at future high energy hadron colliders.  As such it
is vital to have reliable predictions for the production rates both in
the Standard Model (SM) and in the minimal supersymmetric extension of
the Standard Model (MSSM). The two--loop QCD corrections to the main
production process, the gluon fusion mechanism \cite{georgi},  have been
calculated in the SM in Refs.\cite{daws,spira} and later generalized to
the quark contributions in the MSSM \cite{spira,spirasusy}. The
corrections are large and positive, increasing the production rates
significantly. 

The MSSM requires the introduction of two Higgs doublets leading, after
spontaneous symmetry breaking, to two neutral CP--even ($h$ and $H$), a
neutral CP--odd ($A$) and two charged ($H^\pm$) Higgs particles
\cite{susyrev}. While in the SM the dominant contribution to Higgs boson
production in the gluon fusion mechanism originates from top and, to a
lesser extent, bottom quark loops, in the MSSM there are additional
contributions to the production of the CP--even Higgs bosons from scalar
squark loops.  These contributions can be neglected for very heavy
squarks. However, many supergravity--inspired models predict squark (in
particular stop and sbottom squark) masses significantly below 1 TeV
\cite{sugra}. In this case, squark loop contributions to the
Higgs--gluon couplings can be of the same order, or even larger, as the
standard quark contributions, as was recently stressed in
Refs.\cite{twophot}. 

In this letter, we present the ${\cal O}(\alpha_s^3)$  QCD corrections
to the cross sections $\sigma(pp \to \hH + X)$ of the fusion processes
for the neutral CP--even Higgs particles $\hH = h,H$ 
\begin{equation}
gg \to \hH (g)~~~~\mbox{and}~~~~gq \to \hH q,~~ q\bar q \to \hH g
\quad .
\label{eq:proc}
\end{equation}
Because of CP invariance, squark loops do not contribute to the
production of the CP--odd Higgs boson in lowest order. The QCD
corrections from squark loops are evaluated in the heavy squark limit,
where the calculation can be simplified by extending the lowest--order
low--energy theorems \cite{spira,lowen} to two loops. This limit should
be a very good approximation \cite{spira} for the production of Higgs
particles with masses smaller than twice the squark masses. Given the
experimental bounds on the squark masses \cite{bound}, this is fully
justified in the case of the lightest Higgs boson $h$, which is
constrained to be lighter than $\sim 130$ GeV in the MSSM; for the
heavier CP--even Higgs boson $H$, this approximation is valid for masses
smaller than a few hundred GeV. For simplicity, we will restrict
ourselves to the case of degenerate squarks where mixing effects are
absent (in the absence of gluino--exchange, the results can be trivially
generalized to include mixing). Also in this case, scalar squarks will
not contribute to the production of the CP--odd Higgs boson $A$ at
next--to--leading order. 

To lowest order, the cross sections for CP--even Higgs boson production
at proton colliders are given by 
\beq
\sigma_{LO}(pp \to \hH + X) = \sigma_0^{\hH} \tau_{\hH} \frac{d{\cal L}^{gg}}
{d\tau_{\hH}}\quad ,
\eeq
with $d{\cal L}^{gg}/d\tau_{\hH}$ the gluon luminosity at
$\tau_{\hH}=M_{\hH}^2/s$ and $s$ is the total c.m.~energy. The parton
cross sections are built up from  heavy quark and squark amplitudes, 
\beq
\sigma_0^{\hH} = \frac{G_F\alpha_s^2}{128\sqrt{2}\pi} \left| \sum_Q g_Q^{\hH}
A_Q^{\hH} (\tau_Q) + \sum_{\squ} g_{\squ}^{\hH} A_{\squ}^{\hH} 
(\tau_{\squ} ) \right| ,
\eeq
where the sums run over $t,b$ quarks and the left-- and right--handed
squarks $\squ_L, \squ_R$, which in the absence of mixing are identical
to the mass eigenstates. The form factors, with the scaling variables
$\tau_{Q/ \squ}\equiv 4m_{Q/ \squ}^2/M_{\hH}^2$, can be expressed as 
\beqn
A_Q(\tau_Q) & = & \tau_Q \left[ 1+(1-\tau_Q) f(\tau_Q) \right] \\
A_{\squ}(\tau_{\squ}) & = & -\frac{1}{2}\tau_{\squ} \left[ 1-\tau_{\squ}
f(\tau_{\squ}) \right] \ ,
\eeqn
using the scalar triangle integral
\beq
f(\tau) = \left\{ \begin{array}{ll}
\displaystyle \arcsin^2 \biggl(\frac{1}{\sqrt{\tau}}
\biggr) & \tau \ge 1 \\
\displaystyle - \frac{1}{4} \left[ \log \frac{1+\sqrt{1-\tau}}
{1-\sqrt{1-\tau}} - i\pi \right]^2 & \tau < 1
\end{array} \right. \ .
\eeq
The normalized scalar  quark and squark couplings to the CP--even Higgs
bosons, $g^{\hH}_{Q,{\squ}}$, can be found in Refs.\cite{spira,susyrev}.
In the case where all squarks are taken to be degenerate, only the
contributions proportional to the Yukawa--type couplings of the stop and
sbottom squarks have to be added to the top and bottom quark loop
contributions. The couplings as well as the CP--even Higgs masses are
determined at tree--level by two parameters, which are generally chosen
to be the ratio of the vacuum expectation values of the two Higgs
fields, $\tgb$, and the pseudoscalar Higgs mass, $M_A$. All MSSM Higgs
masses and couplings are calculated using the two--loop renormalization
group improved effective potential \cite{carena}. 

The QCD corrections to the gluon fusion process, eq.(\ref{eq:proc}),
consist of virtual two--loop corrections and one--loop real corrections
due to gluon radiation, as well as contributions from quark--gluon
initial states and quark--antiquark annihilation.  The renormalization
program has been carried out in the $\overline{\rm MS}$ scheme for the
strong coupling constant and the parton densities, while the quark and
squark masses are defined at the poles of their respective propagators.
The result for the cross sections can be cast into the form: 
\beqn
\sigma(pp\to \hH +X) & = & \sigma^{\hH}_0 \left[1+C_{\hH}(\tau_Q, \tau_{\squ})
\frac{\alpha_s}{\pi}
\right] \tau_{\hH} \frac{d{\cal L}^{gg}}{d\tau_{\hH}} \nonumber \\
& & + \Delta \sigma^{\hH}_{gg}
+ \Delta \sigma^{\hH}_{gq} + \Delta \sigma^{\hH}_{q\bar q}
\label{eq:sigstruc}
\quad .
\eeqn
The coefficient $C_{\hH}$ denotes the virtual two--loop corrections
regularized by the infrared singularities of the real gluon emission.
The terms $\Delta \sigma^{\hH}_{ij}$ ($i,j=g,q$) denote the finite parts
of the real corrections due to gluon radiation and the $gq$ and $q\bar
q$ initial states. The expressions for the $t,b$ quark contribution can
be found in Refs.\citer{daws,spirasusy}. 

The calculation of the QCD corrections has been performed by extending
the low--energy theorems \cite{spira,lowen} to scalar squarks at the
two--loop level.  For a light CP--even Higgs boson, these theorems
relate the matrix elements of the quark and squark contributions to the
Higgs--gluon vertex to the gluon two--point function. Denoting the
matrix element of the squark contribution to the gluon two--point
function by ${\cal M}_{\squ}(gg)$ and the corresponding matrix elements
with an additional light CP--even Higgs boson by ${\cal
M}_{\squ}(gg\hH)$, one has at lowest order \cite{comlow} 
\beq
{\cal M}_{\squ}(gg\hH) = \sum_{\squ} \left(\sqrt{2}G_F \right)^{1/2}
g_{\squ}^{\hH}
m_{\squ}\frac{\partial {\cal M}_{\squ} (gg)}{\partial m_{\squ}}
\ .
\label{eq:let}
\eeq
To extend this relation to higher orders, all quantities have to be
replaced by their bare values; after differentiation, the
renormalization then has to be performed. In the following we consider
only the pure gluon exchange contributions, which are expected to be the
dominant ones; for heavy enough gluinos, the two--loop corrections due
to gluino exchange should be small since they are suppressed by inverse
powers of the gluino mass.  In this case the differentiation with
respect to  the bare squark mass $m_{\squ}^0$ can be rewritten in terms
of the renormalized mass $m_{\squ}$. A finite contribution to the QCD
corrections arises from the anomalous mass dimension $\gamma_{\squ}$
\cite{spira} 
\beq
m_{\squ}^0
\frac{ \partial}{\partial m_{\squ}^0} = \frac{m_{\squ}}{1+\gamma_{\squ}}~
\frac{ \partial}{\partial m_{\squ}}
\quad .
\eeq
The remaining differentiation with respect to the renormalized squark mass of
the gluon two--point function leads to the squark contribution $\beta_{\squ}$
to the QCD
$\beta$ function. The final result for the squark contributions to the $\hH$
coupling to gluons can be expressed in terms of the effective Lagrangian
\beq
{\cal L}_{eff}^{\squ} = \left( \sqrt{2} G_F \right)^{1/2} \sum_{\squ}
\frac{g_{\squ}^{\hH}}{4}
\frac{\beta_{\squ}(\alpha_s)/\alpha_s}{1+\gamma_{\squ}(\alpha_s)} 
G^{a\mu\nu} G^a_{\mu\nu} \hH .
\eeq
The QCD corrections are then fully determined by the anomalous mass dimension 
of the squarks \cite{sugra,susybet}
\beq
\gamma_{\squ} = \frac{4}{3}~\frac{\alpha_s}{\pi} + {\cal O}(\alpha_s^2)
\eeq
and the squark contribution to the QCD $\beta$ function \cite{betafun}
\beq
\frac{\beta_{\squ}(\alpha_s)}{\alpha_s} = \frac{\alpha_s}{12\pi} \left[
1+ \frac{11}{2} \frac{\alpha_s}{\pi} \right] + {\cal O} (\alpha_s^3)
\ ,
\eeq
resulting in a final rescaling of the lowest--order Lagrangian by a
factor $1+ 25 \alpha_s/6\pi$ at next--to--leading order, compared to a
rescaling factor $1+ 11 \alpha_s/4\pi$ for the quark contribution.
Starting from the Lagrangian eq.(10), the effective QCD corrections due
to real gluon emission and the $gq/q\bar{q}$ initial states have to be
added. These corrections are identical to the corresponding corrections
to quark loops \cite{spira} in the heavy quark limit. 

The QCD corrected squark loop amplitudes have to be added coherently to
the corrected $t,b$ loop amplitudes, whose full mass dependence
is known. To obtain a more reliable prediction for the total
cross sections, the resulting amplitudes for the squark contributions
have been normalized to the lowest--order amplitude in the limit of large
squark masses. These ratios are then multiplied by the lowest--order
amplitude including the full squark mass dependence. The heavy squark
limit is then expected to be a very good approximation for Higgs masses
below the ${\squ} {\squ}^*$ threshold, as in the corresponding case of
top quark contributions \cite{spira}. 

The final results for the partonic cross sections defined in
eq.(\ref{eq:sigstruc}) can be expressed as 
\beqn
C_{\hH}(\tau_Q, \tau_{\squ}) & = & \pi^{2}+c^{\hH}_1(\tau_Q, \tau_{\squ})
+\frac{33-2N_{F}}{6}
\log\frac{\mu^{2}}{M_{\hH}^{2}} \nonumber \\
\Delta \sigma^{\hH}_{gg} & = & \int_{\tau_{\hH}}^{1} d\tau \frac{d{\cal
L}^{gg}}{d\tau} \frac{\alpha_{s}}{\pi} \sigma^{\hH}_{0} \left\{ 
-\hat{\tau} P_{gg} (\hat{\tau}) \log \frac{M^{2}}{\tau s}
\right. \nonumber \\
& & \hspace{3cm} + d^{\hH}_{gg}(\hat\tau, \tau_Q, \tau_{\squ})
\nonumber \\
\label{eq:sigres}
& & + 6 \left.  \left[ 1+\hat\tau^4 + (1-\hat\tau)^4 \right] \left( \frac{
\log(1-\hat{\tau})}{1-\hat{\tau}} \right)_{+} \right\} \nonumber \\
\Delta \sigma^{\hH}_{gq} & = & \int_{\tau_{\hH}}^{1} d\tau
\sum_{q,\bar{q}} \frac{d{\cal L}^{gq}}{d\tau} \frac{\alpha_{s}}{\pi}
\sigma^{\hH}_{0} \left\{ \phantom{\frac{1}{2}}\!\!\! d^{\hH}_{gq}
(\hat\tau, \tau_Q, \tau_{\squ}) 
\right. \nonumber \\
& & \left.  -\frac{\hat{\tau}}{2}
P_{gq}(\hat{\tau}) \left[ \log\frac{M^{2}}{\tau s} - 2\log(1-\hat{\tau})
\right] \right\} \\
\Delta \sigma^{\hH}_{q\bar{q}} & = & \int_{\tau_{\hH}}^{1} d\tau
\sum_{q} \frac{d{\cal L}^{q\bar{q}}}{d\tau}~\frac{\alpha_{s}}{\pi}
\sigma^{\hH}_{0} \ d^{\hH}_{q\bar q}(\hat\tau, \tau_Q, \tau_{\squ}) \nonumber
\eeqn
with $\hat\tau = \tau_{\hH}/\tau$ and $N_F$ being the number of light
flavors contributing to the evolution of $\alpha_s$ and the parton
densities. The renormalization scale $\mu$ enters the lowest--order
expression $\sigma_0^{\hH}$ as the scale of the strong coupling
$\alpha_s=\alpha_s(\mu^2)$. $P_{gg}, P_{gq}$ denote the
Altarelli--Parisi splitting functions \cite{altpar};  $M$ is the
factorization scale at which the parton luminosities are evaluated.
$F_+$ is the usual $+$ distribution, $F(\hat\tau)_+ = F(\hat\tau) -
\delta(1-\hat\tau) \int_0^1 dx F(x)$. 

The contributions to the coefficients $c_1^{\hH}$ and $d_{ij}^{\hH}$
appearing in eq.(13) from squarks, in the heavy squark limit without
$t,b$ loops, are given by 
\beqn
c_1^{\hH}\to \frac{25}{3}~~, ~~d_{gg}^{\hH}\to -\frac{11}{2}
(1-\hat\tau)^3 \nonumber \\
d_{gq}^{\hH}\to -1 + 2\hat\tau - \frac{1}{3}\hat\tau^2~~,~~ 
d_{q\bar q}^{\hH}\to \frac{32}{27} (1-\hat\tau)^3
\eeqn

\begin{figure}[htb]
\epsfxsize=8.2cm \epsfbox{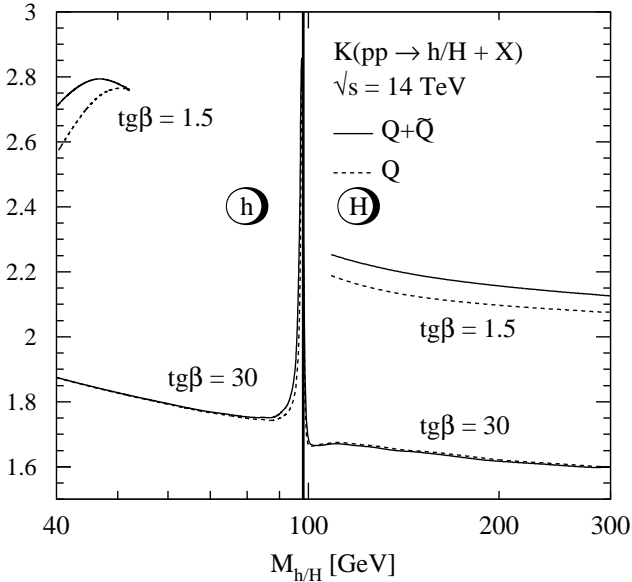}
\caption[]{K factors of the cross sections $\sigma(pp\rightarrow \hH + X)$
for $\tgb = 1.5$ and 30. The solid lines include $t,b$ as well as squark
contributions, the dashed lines include only the $t,b$ contributions.
The common squark mass is chosen to be $m_{\squ}=200$ GeV. We take
$m_b=5$ GeV, $m_t=176$ GeV and use the next--to leading order
$\alpha_s$, fixed by the world average value $\alpha_s(M_Z^2) = 0.118$
\cite{bethke}. The cross sections are convoluted with next--to--leading
order GRV parton densities \cite{GRV}. The renormalization scale $\mu$
and the factorization scale $M$ are identified with the Higgs masses.} 
\end{figure}

In Fig.1, we present the K factors for  the QCD corrections to the
production of the CP--even MSSM Higgs bosons as functions of the ${\hH}$
masses for the LHC at a c.m.~energy $\sqrt{s} = 14$ TeV with (solid
lines) and without (dashed lines) the squark contributions. The K
factors are defined as the ratios of the QCD corrected and lowest--order
cross sections, using next--to--leading order $\alpha_s$ and parton
densities in both terms. A common value $m_{\squ}= 200$ GeV has been
used for the left-- and right--handed stop and sbottom squark masses.
This value is identified with the SUSY scale of the MSSM couplings 
and Higgs masses, leading to a rather low upper limit on the lightest 
Higgs mass $M_h$ for a given value of $\tgb$. 

The QCD corrections enhance the cross sections by a factor between 1.6
and 2.8, if the lowest--order cross sections are evolved with
next--to--leading order $\alpha_s$ and parton densities. If the
lowest--order cross sections are convoluted with lowest order $\alpha_s$
and parton densities, the K factors are reduced to a level between 1 and 2.
It can be inferred from Fig.1 that the inclusion of squark loops in the
production of both CP--even Higgs particles $h$ and $H$ does not
substantially modify the K factors compared to the case where squark
loops are absent. The ${\cal O} (10\%)$ discrepancy between the two
factors for small $\tgb$ (where both the top and the stop contributions
are dominant and can be approximated by their heavy mass limits) is
mainly due to the difference between the contribution of quarks and
squarks to the effective Lagrangian ($c_1^{\hH}=\frac{25}{3}$ for $\squ$
and $c_1^{\hH}= \frac{11}{2}$ for $Q$ loops). 

We have verified that the K factors do not depend significantly on the
squark mass which enters the MSSM couplings and lowest--order cross
sections in our analysis. In fact, in the extreme situation where one of
the $\tilde{t}, \tilde{b}$ squark eigenstates is  relatively light while
the other squarks are heavy and decouple (as is the case for large
squark mixing), the K factors are almost the same as in Fig.1.
Therefore, while it substantially changes the Higgs--squark couplings
and hence the production cross sections, mixing in the stop or sbottom
sectors should have a rather modest impact on the K factors. 

Thus, to a good approximation, the effect of the squark loops in the
gluon fusion mechanism is quantitatively determined by the lowest--order
cross section (including squark loop contributions), multiplied by the
known K factors when only the $t,b$ quark contributions
\cite{spira,spirasusy} are taken into account. 

In Fig.2, we illustrate the effect of including the squark loops and the
QCD corrections to the production rate of the lightest CP--even Higgs
particle.  The ratio of the QCD corrected cross sections with and
without squark loops is shown as a function of the common squark mass
for three values of $\tgb = 1.5,3$ and 30, with the pseudoscalar Higgs
mass fixed to $M_A=100$ GeV. The squark contributions increase the
production cross sections significantly for squark masses below about
500 GeV, especially for small and moderate values of $\tgb$; for higher
masses $m_{\tilde{Q}}$, the squarks decouple from the amplitude. This
large effect can be understood by recalling that for these values of
$\tgb$, the top and stop contributions dominate and the inclusion of the
stop loops leads, in the heavy top and squark limit, to the enhancement
of the lowest--order production cross section by an amount 
\beqn
\frac{\sigma_{t+\tilde{t}}}{\sigma_t} = \left(1 + \frac{1}{2}\frac{m_t^2}
{m^2_{\tilde{t}}} \right)^2
\eeqn
where $\sigma_{t+\tilde{t}}$ ($\sigma_t$) denotes the cross section
including (without) stop loops. Thus for stop masses $m_{\tilde{t}}$ of
the order of the top mass $m_t$ the cross sections are significantly
enhanced by including stop loops. For large values of $\tgb$, because
the Higgs couplings to (s)top (s)quarks are strongly suppressed (except
for $h$ in the decoupling limit), and the contribution of the bottom
loop is enhanced by large logarithms compared to the sbottom loop, the
production cross section is significantly affected only by light squark
contributions, while they become negligible 
 for sbottom masses of the order of 200 GeV.  
\begin{figure}[htb]
\epsfxsize=7.2cm \epsfbox{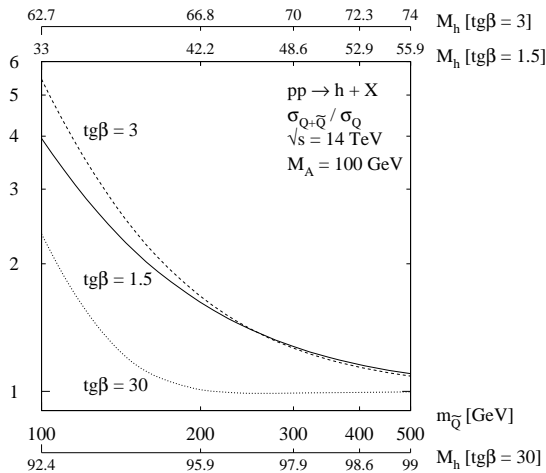}
\caption[]{Ratio of the QCD corrected cross sections $\sigma(pp
\rightarrow h + X)$ with and without squark loops for three values of
$\tgb = 1.5,3, 30$, and for $M_A=100$ GeV. The secondary axes present the
corresponding Higgs masses $M_h$. The quark masses, $\alpha_s$ and the
parton densities are as in Fig.1.} 
\end{figure}
     
Light squarks, with masses $m_{\squ} < 300$ GeV, can have a large impact
on the production cross sections of the CP--even MSSM Higgs bosons at
hadron colliders. We have presented the QCD corrections to the gluon
fusion processes $pp\to gg \to h/H$ including squark loops, in the heavy
squark limit, which should well describe the full corrections including
mass effects, at least in the range where the Higgs masses are
smaller than twice the squark masses. The $t$ and $b$ quark mass
dependence has been included exactly. To a good approximation the K
factors are the same as the corresponding K factors when only top and
bottom contributions are included. Since they increase the production
cross sections significantly, the QCD corrections must be taken into
account. 

\smallskip

Discussions with Jon Bagger and Manuel Drees about SUSY beta functions
and Jim Wells about Ref.\cite{twophot} are gratefully acknowledged. 
S.D.~ is supported by the US Dept. of Energy under contract
DE-AC02-76CH00016. A.D.~is supported by the DFG, Bonn.

\end{narrowtext}

\begin{references}  

\bibitem{georgi} H.M.~Georgi, S.L.~Glashow,
 M.E.~Machacek and D.V. Nanopoulos,
{\it Phys.~Rev.~Lett.}~{\bf 40} (1978) 692.

\bibitem{daws}
S.~Dawson, {\it Nucl. Phys.} {\bf B359} (1991) 283; 
A.~Djouadi, M.~Spira, and P.M.~Zerwas, {\it Phys. Lett.} {\bf B264} (1991) 440;
D.~Graudenz, M.~Spira, and P.M.~Zerwas, {\it Phys. Rev. Lett.} {\bf 70} (1993)
1372;
M.~Spira, DEST T--95--05, hep-ph/9510347.

\bibitem{spira}
M.~Spira, A.~Djouadi, D.~Graudenz,
and P.M.~Zerwas, {\it Nucl. Phys.} {\bf B453} (1995) 17.

\bibitem{spirasusy}
M.~Spira, A.~Djouadi, D.~Graudenz,
and P.M.~Zerwas, {\it Phys. Lett.} {\bf B318} (1993) 347.

\bibitem{susyrev}
{\it For a review on the MSSM see e.g.}
J.~Gunion, H.E. Haber, G.~Kane and S.~Dawson, {\it The
Higgs Hunter's Guide} (Addison-Wesley, Redwood
City,  1990);
J.~Bagger, TASI-91 (World Scientific, Singapore, 1992).

\bibitem{sugra} {\it See for instance,} V.~Barger, M.~Berger, and P.~Ohmann,
{\it Phys. Rev.} {\bf D47} (1994) 1093.

\bibitem{twophot} 
 G.~Kane, G.~Kribs, S.~Martin, and J.~Wells,
{\it Phys. Rev.} {\bf D53} (1996) 213; 
B.~Kileng, P.~Osland and P.~Pandita, hep-ph/9601284. 

\bibitem{lowen}
J.~Ellis, M.K.~Gaillard and D.V.~Nanopoulos, {\it Nucl. Phys.} {\bf B106}
(1976) 292;
A.~Vainshtein, M.~Voloshin, V.~Zakharov and
M.~Shifman, {\it Sov. J. Nucl. Phys.} {\bf 30}
(1979) 711; 
B.~Kniehl and M.~Spira, {\it Z.~Phys.}
{\bf C69} (1995) 77.
 
\bibitem{bound} F.~Abe {\it et al.}, CDF Collaboration, {\it Phys.~Rev.~Lett.}
{\bf 75} (1995) 613;
S.~Abachi {\it et al.}, D0 Collaboration, {\it Phys. Rev. Lett.}
{\bf 75} (1995) 618;
J.J.~Hernandez, Proceedings, La Thuile Conference 1996 (G.~Bellettini and
M. Greco, eds.), to appear.
 
\bibitem{carena} 
M.~Carena, J.~Espinosa, M.~Quiros, and C.~Wagner,
{\it Phys. Lett.} {\bf B355} (1995) 209.

\bibitem{comlow} Note that eq.(\ref{eq:let}) is only valid if
four--point interactions between Higgs bosons and squarks
do not contribute.

\bibitem{susybet}
L.~Alvarez--Gaum\'e, J.~Polchinski, and M.~Wise, {\it Nucl. Phys.} {\bf B221}
(1983) 495;
S.~Martin and M.~Vaughn, {\it Phys. Rev.} {\bf D50} (1994) 2282;
J.~Derendinger and C. Savoy, {\it Nucl.~Phys. } {\bf B237} (1984)
307.

\bibitem{betafun}
W.~Caswell, {\it Phys.~Rev.~Lett.}~{\bf 33} (1974) 244;
D.R.T. Jones, {\it Phys. Rev.} {\bf D25} (1982) 581;
M. Einhorn and D.R.T. Jones, {\it Nucl. Phys.} {\bf B196}
(1982) 475. 

\bibitem{altpar} G.~Altarelli and G.~Parisi, {\it Nucl.~Phys.}~{\bf B126}
(1977) 298.

\bibitem{bethke} S.~Bethke, Proceedings QCD 94, Montpellier 1994; Report
PITHA--94--30.

\bibitem{GRV} M.~Gl\"uck, E.~Reya and A.~Vogt, {\it Z.~Phys.}~{\bf C53} (1992)
127.

\end{references}
\end{document}